\documentclass{IOS-Book-Article}
\usepackage{mathptmx}
\usepackage{amssymb}
\usepackage{graphicx}
\usepackage{url}
\usepackage{enumitem}
\usepackage{CJKutf8}
\usepackage{color}
\usepackage{pinyin}

\def\hb{\hbox to 10.7 cm{}}

\begin{document}

\pagestyle{headings}
\def\thepage{}

\begin{frontmatter}              

\title{On Constructing a Knowledge Base of Chinese Criminal Cases}


\author[A]{\fnms{Xiaohan} \snm{Wu}
\thanks{Corresponding Author, E-mail:
x2510@columbia.edu}},
\author[A] {\fnms{Benjamin L.} \snm{Liebman}},
\author[B] {\fnms{Rachel E.} \snm{Stern}}, 
\author[C]{\fnms{Margaret E.} \snm{Roberts}},
and
\author[C]{\fnms{Amarnath} \snm{Gupta}%
\thanks{Corresponding Author, E-mail:
a1gupta@ucsd.edu}}

\address[A]{Columbia Law School, USA}
\address[B]{University of California Berkeley, USA}
\address[C]{University of California San Diego, USA}
\begin{abstract}
We are developing a knowledge base over Chinese judicial decision documents to facilitate landscape analyses of Chinese Criminal Cases. We view judicial decision documents as a mixed-granularity semi-structured text where different levels of the text carry different semantic constructs and entailments. We use a combination of context-sensitive grammar, dependency parsing and discourse analysis to extract a formal and interpretable representation of these documents. Our knowledge base is developed by constructing associations between different elements of these documents. The interpretability is contributed in part by our formal representation of the Chinese criminal laws, also as semi-structured documents. The landscape analyses utilizes these two representations and enables a law researcher to ask legal pattern analysis queries.
\end{abstract}

\begin{keyword}
landscape analysis, Chinese criminal cases, Information Extraction,  discourse analysis, context-sensitive grammar,knowledge representation
\end{keyword}
\end{frontmatter}



\section{Introduction}
A legal system and the cases that flow through it is a reflection of the society in which the system operates. The norms and conventions of the society not only decide the way laws are structured but the way everyday jurisprudence is conducted. For example, a legal researcher who is trained in US laws, and fully conversant with American legal practices, may find it surprising that judges tend to give lenient punishments in a battery case if the indictable offense resulted from family conflicts or neighborhood disputes\cite{WuJudicial} -- not because the damages in these cases are less severe, but because victims in these cases are likely to share part of responsibility of the indictable offense in the societal culture and consequently in the legal practice. 

Our long-term goal is to develop a knowledge-based information system that would capture this ``general knowledge'' about a legal universe and the way law is practised in that universe.
We use the term ``general knowledge'' in the following sense. Instead of asking descriptive questions about a single case such as the case summary, the argument structure, whether a law was appropriately applied to a case, how to predict the decision of a case given the prosecution's and defense's arguments etc. for specific cases, we would like the system to provide answers to questions regarding ``what usually happens" in a given scenario and what makes some case exceptional. For example, using the query operations of the system, the researcher should be able to discover that no defense argument is usually presented for drunk driving cases, and in an exceptional situation where there is one, only a leniency in the punishment is requested. Similarly, the the researcher should be able to formulate queries (and get answers) regarding the extent to which the full range of punishments allowed for a specific type of crime is actually meted out. We call these class of questions \textit{legal landscape analyses}, formalized later in the paper. 


\noindent \textbf{Prior Work.} The primary corpus for our study is the Judicial Decision Documents (JDD) available from the Supreme People's Court (SPC) \cite{liebman2017mass}. As Gupta et al \cite{GuptaJurix17} showed, parts of the data, such as the parties to the lawsuit including the plaintiffs and defendants, together with their legal representation, are represented as structurable text, stored in a relational database. They also indexed the JDDs with a legal ontology extended from the SPC-provided classification of case types. However, \cite{GuptaJurix17} did not analyze the unstructured part of the text that contains the legal arguments presented by the counsels as well as the facts found by the court. 

\noindent \textbf{Challenges.} In this paper, we address three primary challenges that go beyond \cite{GuptaJurix17}.
\begin{enumerate}[leftmargin=*]
    \item \textit{Landscape Analysis Framework.} Branting \cite{branting2017data} defines the term ``legal landscape" as a collection level analysis that provides a global characterization of the state of the law relevant to a given set of tasks. In their example, ``a patent landscape is the collection of existing patents relevant to (e.g., representing prior art similar to) a given topic." We take a more operational view of the term. In \cite{GuptaJurix17}, the JDD was represented as a relation (table), whose attributes may be have semi-structured or unstructured values. Landscape analysis treats each column of this table as a collection of features whose distribution patterns must be queried and analyzed. To accomplish this:
    \begin{itemize}[leftmargin=*]
        \item A set of features suitable for landmark analysis must be developed. These features will be different from those designed for ontology-based knowledge summarization \cite{ma2018efficient}, semantics extraction \cite{lin2017}, linguistic analysis \cite{yuan2019linguistic} and so forth.
        \item The feature representation must itself be used for querying and inference.
        \item The features should enable us to determine ``normal'' and ``outlier'' patterns, and yet allow conventional query-answering.
    \end{itemize}
    \item \textit{Information Extraction from Facts.} Linguistic variability is known as a major impediment in analyzing textual data. For our intended task, the feature set for describing the facts and punishments must be reliably extracted by NLP techniques without taking recourse to simplistic methods like just regular expressions (see \cite{dong2018establish} as an example). We show in the paper why this task can be difficult because of the specific nuances of the Chinese language as well as the stylized use of the language of JDDs.
    \item \textit{Answering Analytical Questions.} The ability to formulate and answer landscape analysis questions is the complementary task of feature design and representation. To our knowledge, there is no formal language, query primitives or analytical libraries for specification of landscape analyses. Similarly, while there are many algorithms to analyze feature distributions, to our knowledge, there is no query evaluation or question answering mechanism supported by any knowledge-based legal information system.
\end{enumerate}


\begin{figure}
    \centering
    \includegraphics[width=\textwidth]{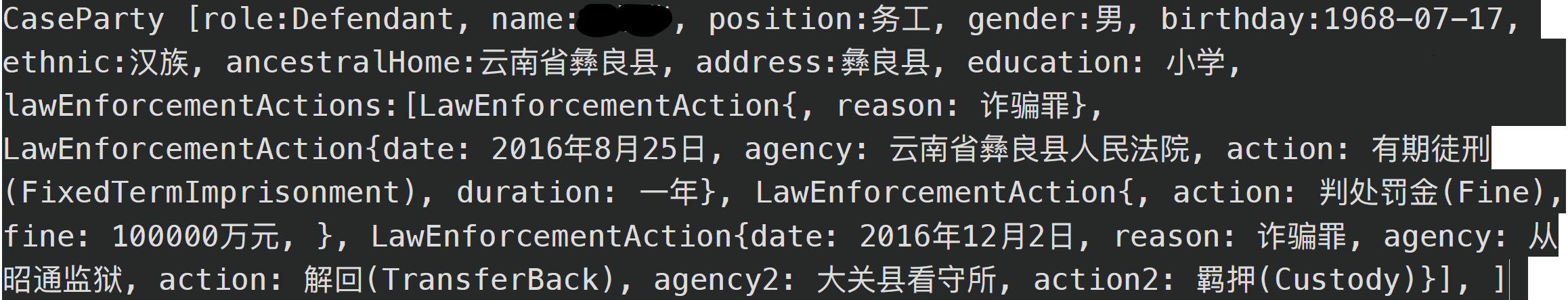}
    \caption{The semi-structured output of a party involved in a case.}
    \label{fig:parties}
\end{figure}
\vspace{-1em}
\section{Landscape Analysis of Legal Documents - A First Formal Model}
\label{sec:landscape}
We model a collection $\mathbf{C}$ of JDDs as a triple $(\mathbf{S, D, M})$ where $\mathbf{S}$ is a heterogeneous relation, $\mathbf{M}$ is a $k$-dimensional matrix and $\mathbf{D}$ is a mapping between elements of $\mathbf{S}$ and the indices of $\mathbf{M}$. Here, a \textit{heterogeneous relation} refers to a relation whose attributes can take different forms of semi-structured values. For example, \texttt{case-type} is a string valued (e.g., `criminal' or `administrative') attribute, while  \texttt{parties} is a complex value as shown in Fig. \ref{fig:parties}. Notice how the parser output includes the criminal history of the defendant under the element \texttt{LawEnforcementActions} containing a hierarchy of subelements like the duration of the defendant's imprisonment. 

The matrix $\mathbf{M}$ is derived from our analysis of the text-valued \texttt{Fact} element. Using parsing methods described in the next section, sentences in the fact can be classified into 8 classes:
\begin{CJK*}{UTF8}{gbsn}
1. 案件由来 -- case background,
2. 原告诉称 -- arguments from plaintiff/prosecutor, 
3. 原告证据 -- evidences provided from plaintiff/prosecutor, 
4. 原告意见 -- requests/opinions from plaintiff/prosecutor, 
5. 被告辩称 -- arguments from defendant, 
6. 被告证据 -- evidences from defendant, 
7. 事实认定 -- reviewed facts from court, and 
8. 认定证据 -- evidences accepted by court.
\end{CJK*}
In a typical JDD document, multiple consecutive sentences may belong to each class.  The sentences in these sections can be further decomposed into an \textit{action schema} given by \texttt{[subject, action, object, action\_modifier]}. For example, the sentence (translated) ``The defendant surrendered himself at police station in Binjiang on Feb.13th, 2017, where he admitted his crime honestly." has the actions:
\begin{verbatim}
[['name of defendant'], 'went to', ['police station in Binjiang'],['voluntarily']] 
[['name of defendant'], 'stated', ['criminal action'],['later',honestly']]
\end{verbatim}
In the sentence 
\begin{CJK*}{UTF8}{gbsn}上述赃物价值共计人民币25920元。(The total value of stolen items is 25,920 yuan.), the system detects the variable \texttt{damage}:
[['25920元']] (25,920 yuan)
\end{CJK*}
A similar representation of the court decision leads to a structure of the punishment issued by the court. For criminal cases \texttt{punishment} is represented by the numeric vector 

\begin{CJK*}{UTF8}{gbsn}
\texttt{\{Exemption(免于刑事处罚), Public Surveillance(管制),Detention(拘役), Fixed-Term Imprisonment(有期徒刑), Probation(缓刑), Fine(罚金), Political Rights Deprivation(剥夺政治权利),  Confiscation(没收), Life Imprisonment(无期徒刑), Death(死刑), 
Political Rights Deprivation For Life(剥夺政治权利终身)\}} 
\end{CJK*}
where \texttt{Death, Exemption, LifeInprisonment, PoliticalRightsDeprivationForLife} are represented in binary code and other vector elements are represented by a quantified ``degree of punishment'' either in terms of time or in terms of monetary value.

\begin{figure}
    \centering
    \includegraphics[width=12cm, height=6cm]{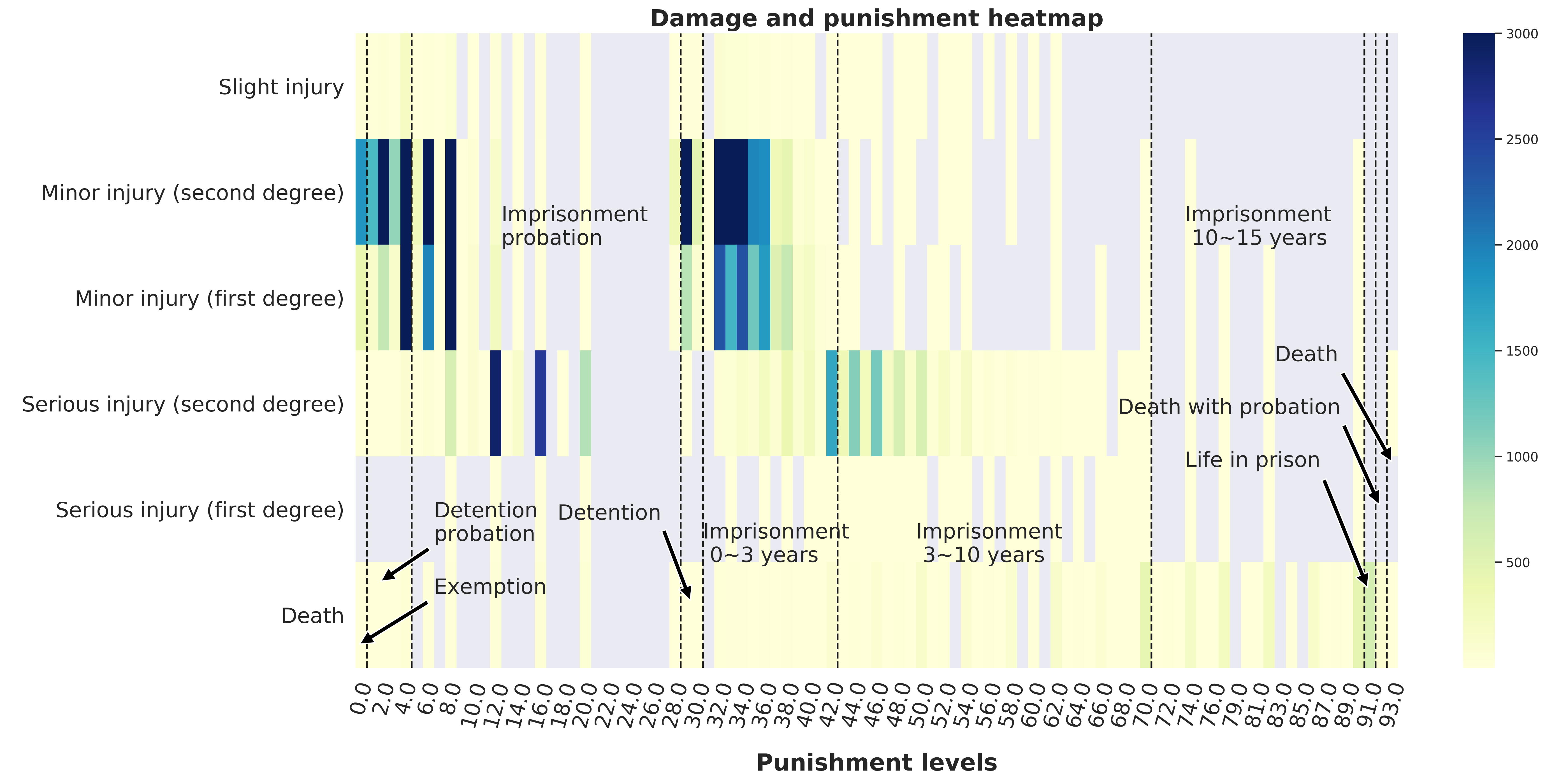}
    \caption{Damage and punishment heat map for assault and battery cases}
    \label{fig:heatmap}
\end{figure}

The representation enables us to represent more than one punishment (e.g., prison time and fine) for a crime. Integrity constraints are applied to ensure that specific combinations of punishments (e.g., \texttt{FixedTermImprisonment} and \texttt{lifeImprisonment}) do not co-occur. We construct the matrix $\mathbf{M}$ as a product \texttt{action} $\times$ \texttt{damage} $\times$ \texttt{punishment-bucket} where a \texttt{punishment-bucket} is a discretized representation of the punishments. A cell of the matrix represents the number of cases that fall in the action-damage-punishment construct.  $\mathbf{M}$ is partitioned by crime type so that theft is considered separately from murder. While this partitioning introduces some inaccuracy for cases where multiple crimes occur, we tolerate the inaccuracy for landscape analyses where the goal is to understand general properties of the distribution. Figure \ref{fig:heatmap} shows a fragment of this matrix as a heatmap.
 Note that the color in this map indicates the number of cases for the corresponding combination. Gray means zero case. 
 The unit for punishment levels is 3 months except for \textit{Exemption, life in prison, death with probation and death penalty}, each of which takes one unit. Figure \ref{fig:heatmap} shows how some combination of damages and punishment are more dense while some other combinations are empty, indicating combinations that although theoretically plausible occur rarely in practice. For example, according to Criminal law article 234, ``whoever intentionally inflicts injury upon another person,causing severe injury to another person, shall be sentenced to fixed-term imprisonment of not less than three years but not more than 10 years''. However, in practice, many assaulters were sentenced to fixed-term imprisonment of less than three years with probation -- indicating judges' discretion in deciding punishments. 

The mapping $\mathbf{D}$ between $\mathbf{S}$ and $\mathbf{M}$, which is used for information retrieval, is a collection of indices. The forward indices serve as a pointer from a schema element like:\\
\texttt{JDD.prosecutorArgument.sentence.actions.drunk-driving} to \\
\texttt{$\mathbf{M}$.traffic-disconduct[3]} where [3] indicates the axis of the matrix where \texttt{drunk-driving} is mapped. Similarly, \texttt{JDD.prosecutorArgument.sentence.drunk-driving.punishment} \\
may map to \texttt{$\mathbf{M}$.traffic-disconduct[3][2]} which is the action-punishment slice of the \texttt{traffic-disconduct} partition of $\mathbf{M}$. In contrast, the reverse index behaves similarly as an inverted index in an information retrieval system where every cell of the matrix is mapped back to a list of case identifiers that populate the cell. Thus, the retrieval function \texttt{getCases($\mathbf{M}$[3][2][4])} will retrieve the drunk driving cases resulting in property damage up to 1000 yuan where a fine was imposed.





\section{Information Extraction}
To extract our analytical primitives, we have developed a parsing strategy for linguistic patterns that are characteristically observed in JDDs. 
The information extraction module assumes that the names of plaintiffs, defendants and their legal counsel are available to the system. 
In the following, we present a method for extracting the ``action'' part from the unstructured \texttt{Facts} of a JDD. The linguistic patterns observed include:

\begin{enumerate}[leftmargin=*]
\item\textit{Mixed-granularity Semi-structured text.} 
As discussed in \cite{dong2018establish, GuptaJurix17}, JDDs are written in fixed forms and are guided by rules given by Supreme People's Court (SPC). \cite{dong2018establish, GuptaJurix17} utilized the semi-structure pattern to parse JDDs.  However, parsed text in these papers is at low-granularity level limiting the accuracy of semantically oriented search and analysis. 
Searching text for light punishment suggested by prosecutor in facts part is likely to return lenient penalties asked by defendants. Here, we define three levels of granularity in JDDs.  
\begin{itemize}[leftmargin=*]
\item Low-granularity data are major section in JDDs such as facts, holding and decision, which can support less elaborate queries in law. 
\item Middle-granularity data are legal components in text describing a case such as a defendant's argument, which supports semantic oriented searches.
\item High-granularity data are structured data that carries semantic information and can be used directly in landscape analysis. 
\end{itemize}
\item \textit{Long flowing sentences.} 
The flowing sentence is a unique sentence pattern in Chinese. It contains so-called \begin{CJK*}{UTF8}{gbsn}链式结构(chain structure)\end{CJK*} --  the relationship between \begin{CJK*}{UTF8}{gbsn}
逗断(\dou4\duan4)
\end{CJK*} was usually indicated by the order of events. Such sentences are very common in JDDs. For example,
\begin{CJK*}{UTF8}{gbsn}
2017年11月1日8时30分，被告人李xx体内藏匿毒品，从缅甸小勐拉走私入境至中国打洛223号界桩，被勐海县公安边防大队打洛封控队查获，执勤人员对其盘问时李xx如实供述体内藏匿毒品，后执勤人员收缴李xx从体内排出的毒品可疑物59坨，净重354克。(At 8:30pm on November 1, 2017, the defendant Li secretly smuggled drugs from Mongla, Burma to the 223 Boundary Pile in China. He was seized by the Luohai County Public Security Frontier Brigade. During the cross-examination by duty officers, Li truthfully confessed that the drug was hidden inside his body. Later, the duty officer collected 59 packets, with a net weight of 354 grams, of suspicious drugs which were excreted from Li's body.) 
\end{CJK*}
The example shows that the semantics of flowing sentences are complex as there is no grammatical restrictions on the amount of information, and there is little or no conjunctions between \begin{CJK*}{UTF8}{gbsn}\dou4\duan4\end{CJK*}. 
Wang \cite{Wang2014Chinese} defined \begin{CJK*}{UTF8}{gbsn}\dou4\duan4 as the basic unit of Chinese text and \dou4\duan4 can be used as the index to specific communication event.\end{CJK*} We use \begin{CJK*}{UTF8}{gbsn}\dou4\duan4\end{CJK*} as the minimum text processing unit for parsing and discourse analysis. \cite{li2005hierarchical} showed that this divide-and-conquer method for long flowing sentences reduces computation and improves parsing accuracy rate.

\item \textit{Action-focused defendant-centered description.} 
The majority of sentences in facts, especially arguments from prosecutor and reviewed facts, are descriptions of actions. \begin{CJK*}{UTF8}{gbsn} As is shown in the flowing sentence example, each \dou4\duan4 is either an action or time/location information. Even if the description is in a passive tone and verbose, the subject of an action is usually the defendant. For example, 被告人已取得被害人家属谅解。(The defendant has already obtained the victim's families' forgiveness.) is much more commonly used than 被害人家属已经谅解了被告人。(The victim's family has already forgiven the defendant.)
\end{CJK*} 
\end{enumerate}
\noindent \textbf{Extracting action triggers.}
Verbs have been used as triggers in open information extraction \cite{fader2011identifying, silveira2016designing} and news events extraction \cite{rusu2014unsupervised}.  These relation patterns, however, is only applicable to English text. Open information extraction research in Chinese is still relatively inadequate\cite{bing2015unsupervised}. In this paper, we focus on extracting central actions where the subjects are mostly the defendant or the police. 
Here we give two rules for trigger verb extraction based on constituency parsing and universal dependency(UD):
\begin{enumerate}[leftmargin=*]
\item\textit{Rule 1.} verbs in paths that originated from ROOT in constituency tree and only contains \textit{\{'IP','VP','VV','VRD'\}} 
\item\textit{Rule 2.} verbs that are \textit{\{'conjunct','clausal complement'\}} dependents of trigger verbs obtained by Rule 1. 
\end{enumerate}
For example, \begin{CJK*}{UTF8}{gbsn}in \dou4\duan4 被告人在15号车厢当面接收张某某发送的手机微信红包(The defendant received Wechat red pockets sent by Zhang in person in car No.15), part-of-speech tagging identified two verbs: 接收(receive) and 发送(send).  The central action in this \dou4\duan4 is, 
[['The defendant'], 'receive', ['wechat red pocket'], ['in person']].
Therefore, the trigger verb is "receive" rather than "send" by \textit{Rule 1}. \end{CJK*}

\noindent \textbf{Extracting elements of actions.}
In addition to action trigger verb, we defined \textit{Subject, Object} and \textit{action\_modifier} in \textit{action schema}.
We extracted these elements based on universal dependencies (a multiliguial generalization of the dependency relationships from the Stanford Dependency parser) of trigger verbs:
\begin{itemize}[leftmargin=*]
\item \textit{Subject} extraction has two rules:  
\textit{Rule 1} extracts nouns that are \textit{'nominal subject'}  of the trigger verb. \textit{Rule 2} inherits \textit{Subject} from the latest \dou4\duan4 if \textit{Rule 1} fails.
\item \textit{Objects} are usually \textit{direct objects} of trigger verbs.  \begin{CJK*}{UTF8}{gbsn} Note that ‘被‘,’将‘ and ’把‘are treated as exceptions. Specifically, if '被' is a \textit{passive auxiliary} dependent of trigger verb, the \textit{object} is the \textit{passive nominal subject} dependent. If '将' and ‘把’ are \textit{auxiliary} dependents of the trigger verb, the \textit{object} is the nominal dependents between '将' or ‘把’ and the trigger verb. 

\item \textit{action\_modifier} are trigger verb's \textit{adverb modifier}.  We also excluded \textit{(遂,并,且,后,但)} because they turned out to be less important in our landscape analysis. 

\end{CJK*} 
\end{itemize}
We verified all rules manually on randomly selected sentences from JDDs and implemented trigger verb extraction with Stanford CoreNLP\cite{manning-EtAl:2014:P14-5}. The extraction accuracy depends on the parsing accuracy of CoreNLP. To reduce errors in trigger verbs, we removed all trigger verbs that appear once and developed an importance score for actions in Section \ref{sec:answering}.

\noindent \textbf{Extract damages, criminal charges,  convicted crime charges and punishments.}
\begin{CJK*}{UTF8}{gbsn} Appraisal agencies evaluate the damages in terms of monetary values or level of injuries. We extract monetary damages by applying named entity recognition(NER). 
There are five injury levels in Chinese legal system - Second degree serious injury, First degree serious injury, Second degree minor injury, First degree minor injury and slight injury. Since the injury levels are fixed and finite, we extracted human health damages by keyword matching.\end{CJK*} 

The first \dou4\duan4 of prosecutor's argument is always criminal charges against defendants and the first \dou4\duan4 in punishment is convicted crimes for corresponding punishments.  We use regular expression to extract the name of crimes in those \dou4\duan4 and convert extracted crime names to standard names to eliminate variations. Currently, there are 469 crime names according to the most recent criminal law amendment.

We use regular expressions to match and extract the punishment as the decision part is highly structured in criminal JDDs. We decided the keywords and extraction details according to the principal and supplementary punishments in Chinese criminal law Article 33. 
If the criminal is a recidivist on probation or commits multiple crimes in one case, judges decide the punishment for each crime separately, and combine the sentences according to Chinese criminal law Articles 69-71. Therefore, punishments mentioned in landscape analysis are penalties for individual convicted crime.


\section{Answering Analytical Questions}
\label{sec:answering}
Analytical queries are queries that combine selection predicates against the heterogeneous relations, as well as analysis operation the distribution matrix. We discuss the formulation as well as the answering of these queries through a set of examples.

\noindent \textbf{Question 1.}  Consider the pair of queries: (a) What is the distribution of punishments for cases where there is a defense argument versus where there is not, conditioned by the damage caused by the crime? (b) Does this distribution depend on whether the defendant received the victims'(or victim families') forgiveness?
    
\begin{figure}
    \vspace{-1em}
    \centering
    \includegraphics[width=12cm, height=5.5cm]{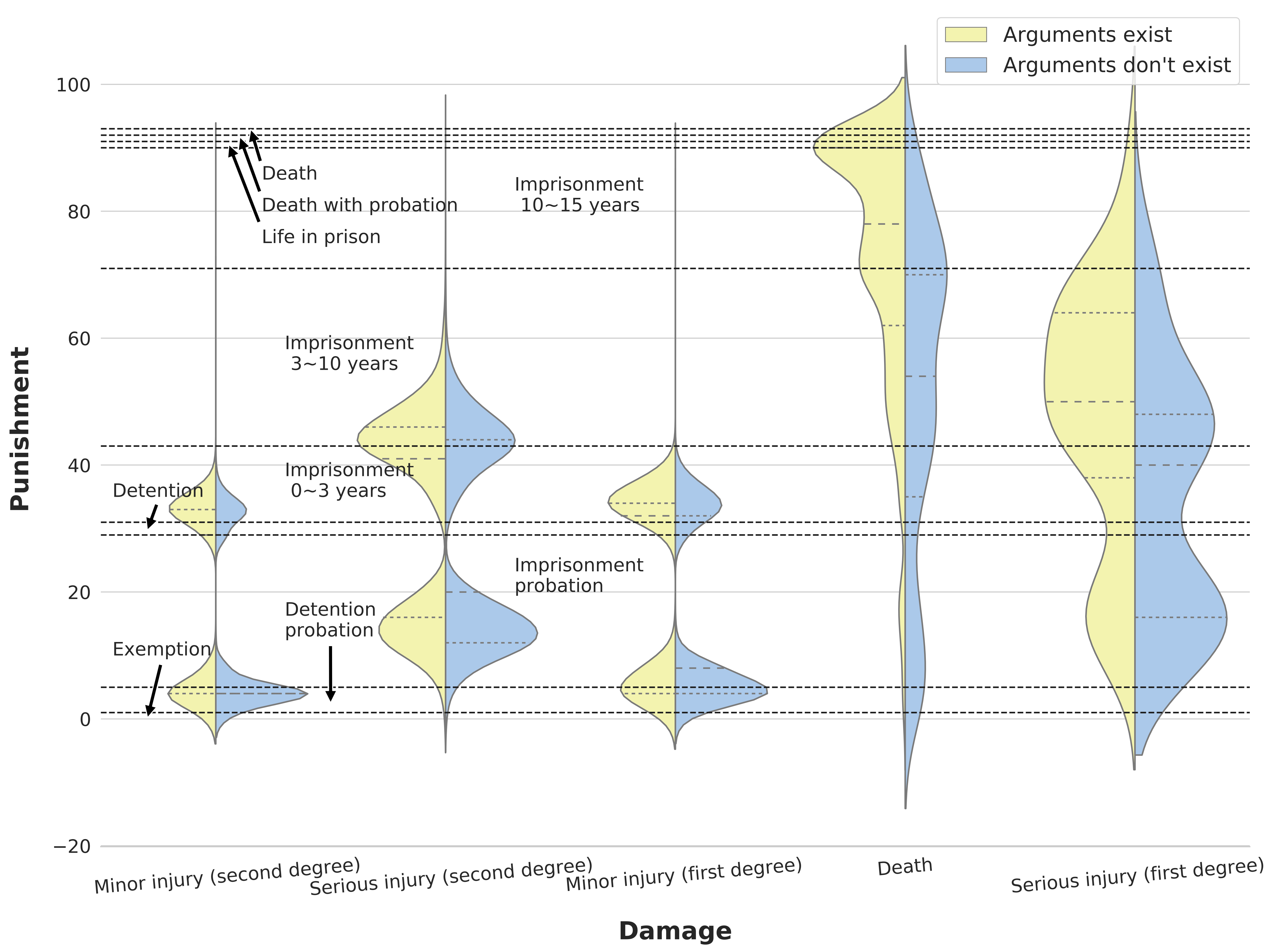}
    \caption{Probability density of punishment levels for battery cases with/without defense argument}
    \label{fig:distribution1}
    \end{figure}

    The query processing steps for part (a) are:
    \begin{enumerate}[leftmargin=*,label=(\roman*)]
        \item $C1$ = get case IDs from cases where the defense argument is not null;
        \item $C2$ = get case IDs from cases where the defense argument is null
        \item $D1$ = list all damages from cases in $C1$
        \item For each damage $d$ in $D1$, 
        \begin{itemize}
            \item find all cells in the matrix whose reverse index intersect with $C1$
            \item extract the non-empty punishment vector for these cells
        \end{itemize}
        \item repeat steps (iii) and (iv) for $C2$
    \end{enumerate}
    For part (a), We identified 42,806 battery cases where there is a defense argument and 86,896 cases where there is not. Most defense arguments made in courts seek lenient penalties. Defense of innocence is very rare. In Figure \ref{fig:distribution1}, the yellow part is probability density of punishments for cases where defense arguments exist while blue part is for cases where defense arguments don't exist. The victim received \textit{Minor injury} or \textit{Serious injury(second degree)} in most battery cases.  The figure shows that defense arguments have little beneficial effect to defendant. For cases involving \textit{Death} or \textit{Serious injury(First degree)}, defendants tend to receive harsh punishments even if they have defense arguments.
    
\begin{figure}
    \vspace{-1em}
    \centering
    \includegraphics[width=12cm, height=5.5cm]{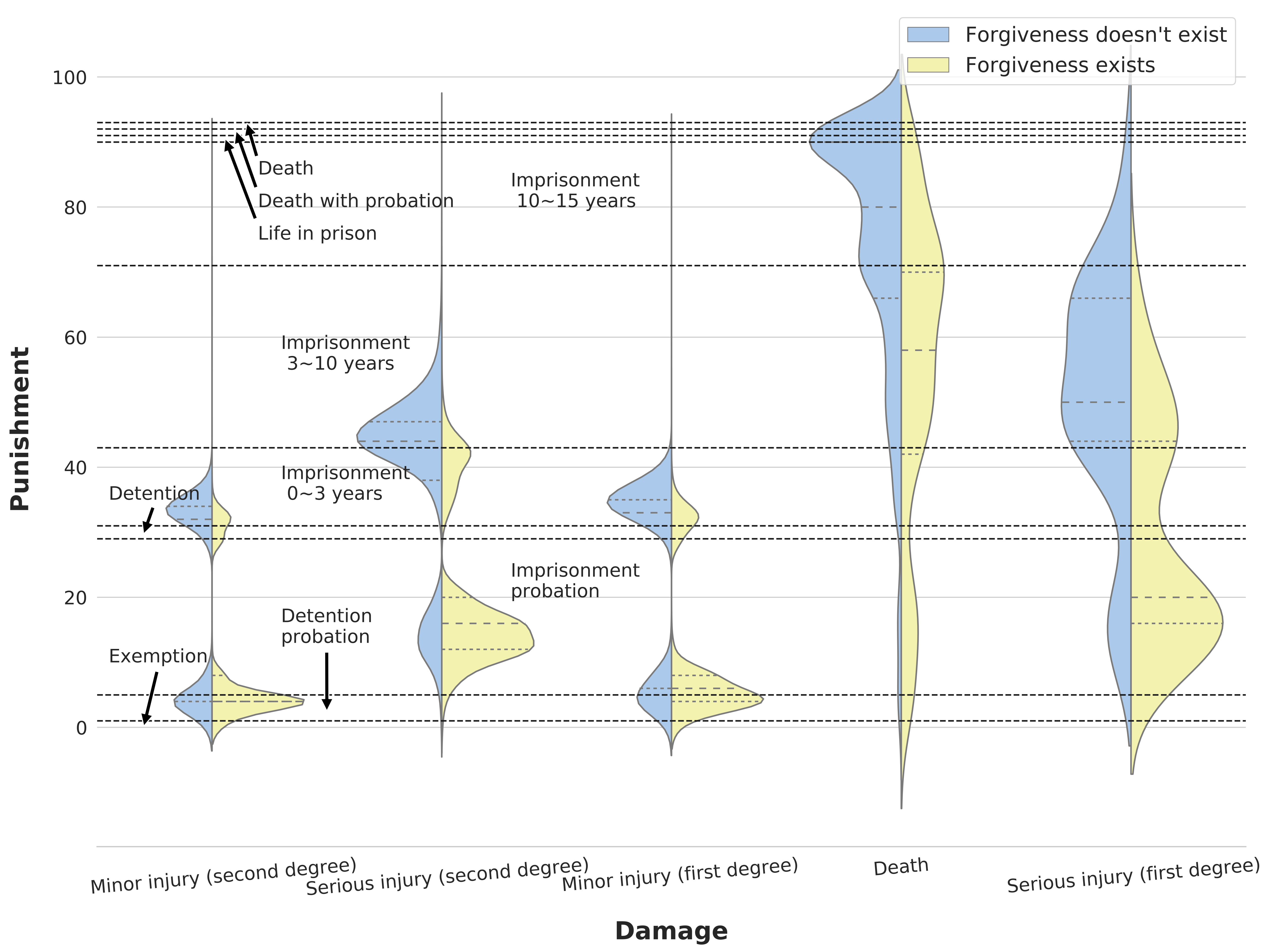}
    \caption{Probability density of punishment levels for battery cases with/without victims' forgiveness}
    \label{fig:distribution2}
    \end{figure}

    For part (b), we define $C1'$ as  a subset of cases where the action includes a lemmatized version of the term ``forgiveness" with positive \textit{action\_modifier} and $C2'$ where the cases do not. The punishment distributions for these two cases are then be presented as output. 
    We found 75655 battery cases where the victim forgave the defendant and 60627 cases where the victim didn't. In Figure \ref{fig:distribution2},the yellow part is probability density of punishments for cases where forgiveness exist while blue part is for cases where forgiveness don't exist. Evidently, judges tend to give lenient punishments to defendants who received forgiveness regardless of the damage severity. 
    
\noindent \textbf{Question 2.}    (a) What punishments are rare for crime type $X$ and under what circumstances are they given? Here, we specify a ``circumstance'' as a combination of crime types, actions and damages. (b) Find the distribution of circumstances for which the punishment is ``exemption''.
    The steps of query evaluation for part (a) are:
    \begin{enumerate}[leftmargin=*,label=(\roman*)]
        \item $P$ = getMarginals($\mathbf{M}.X$, 'punishment')
        \item $cutOff$ = findElbow(sort($P$, 'descending'))
        \item $P'$ = getAxisValues($P.punishment$ where $P.Value > cutOff$)
    \end{enumerate}
    Here, the getMarginals() function computes the marginals of a matrix for the column specified in the argument. Hence, $P$ represents the histogram of all punishments for all combinations of actions and damages. Note that here we count each convicted crime once in getMarginals() function, although each convicted crime may associate with multiple damages and actions. Lines (ii) and (iii) finds the punishments which are ``rare'' by computing the tail of the reverse-sorted (by the value of the histogram) punishment histogram. The getAxisValues() function is a matrix operation that selects a subset of the values of a specified axis that satisfies a given predicate. For part (b), the punishment is fixed as ``exemption''. Thus the query evaluation steps are:
    \begin{enumerate}[leftmargin=*,label=(\roman*)]
        \item $C$ = getMarginals($\mathbf{M}.X$, `punishment'=`exemption')
        \item $C'$ = sort($C$, `descending'))
        \item $C''$ = top-k($C'$, 20)
    \end{enumerate}    
        \begin{figure}
    \centering
    \includegraphics[width=10cm, height=4cm]{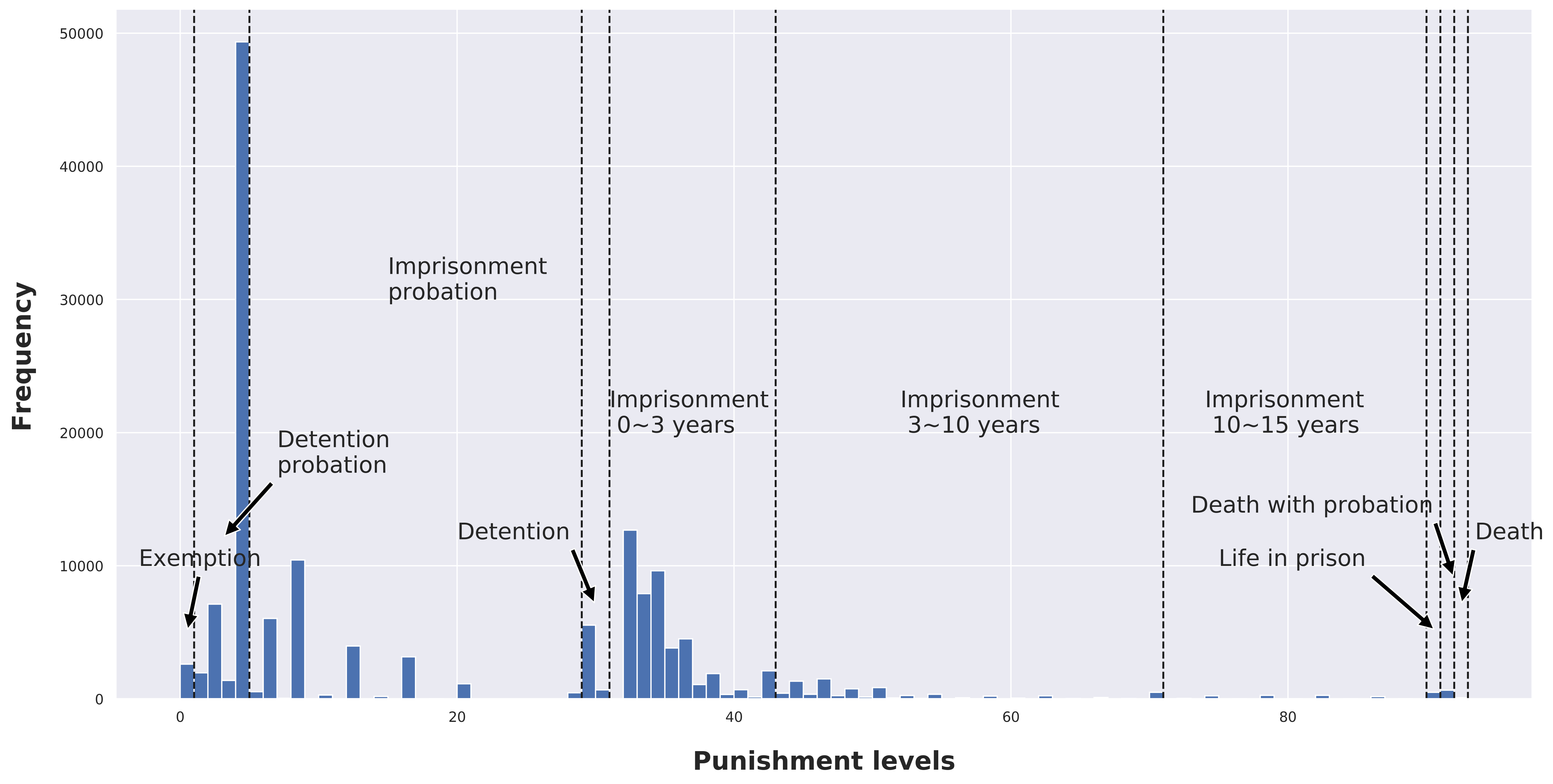}
    \caption{Distribution of punishments for battery cases}
    \label{fig:distribution3}
    \end{figure}
    \begin{figure}
    \vspace{-1em}
    \centering
    \includegraphics[width=10cm, height=4.5cm]{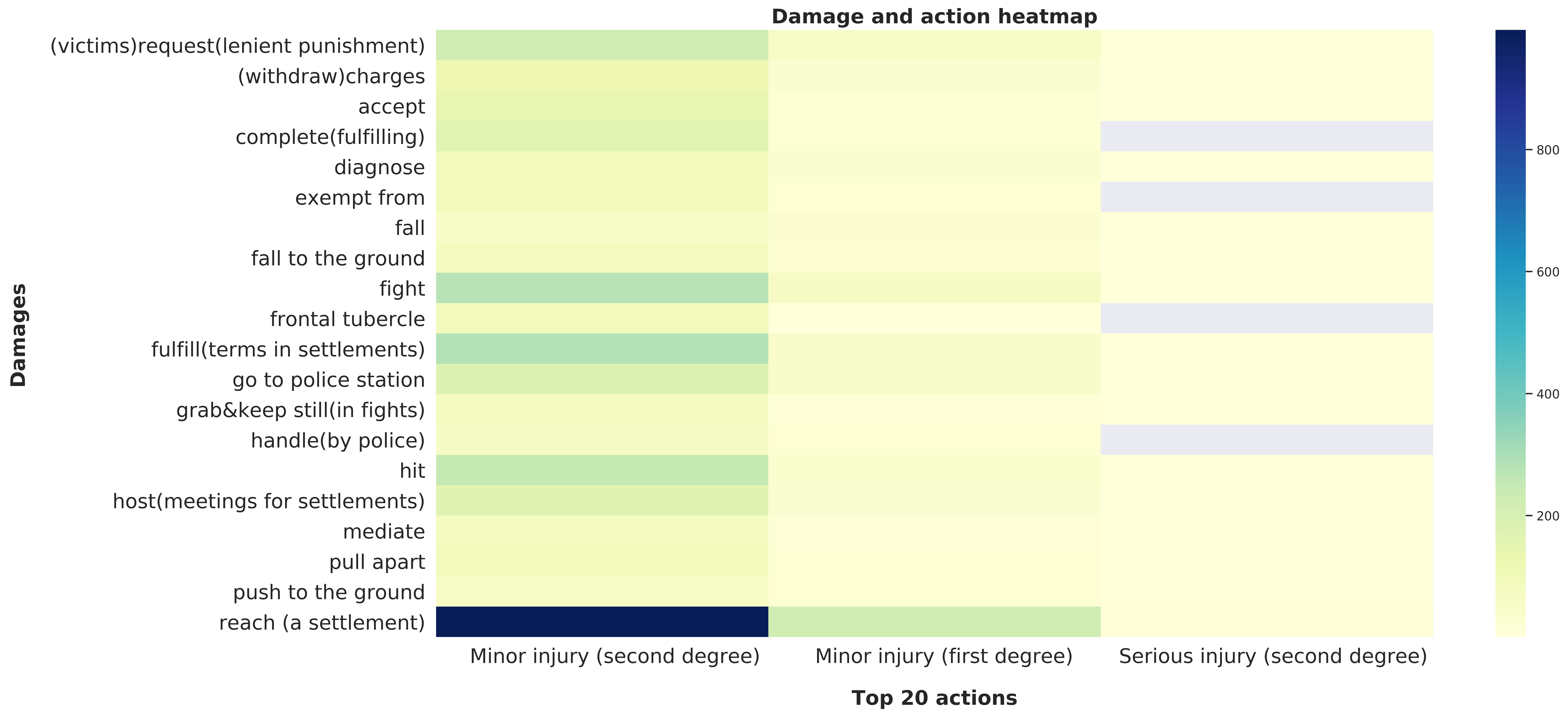}
    \caption{Heat map of damage and top 20 actions in battery cases}
    \label{fig:heatmap2}
    \end{figure}
    Notice the $C$ is a 2D histogram with axes action and damage. $C'$ sorts it descending by value. $C''$ returns a fraction of $C'$ that only contains 20 most important actions defined by user or importance measurement functions. It returns the 20 most frequent action-damage pairs by default. We set \texttt{case type = ``battery''}.  For part (a), we obtained 80 punishment levels . Figure \ref{fig:distribution3} is the histogram of all punishment levels for battery cases. We found two types of rare punishments in $P'$ -- punishments that are extremely lenient or harsh and punishments where the measurement units are not a year, half a year or a quarter.  For part (b), we obtained 2,181 battery cases where defendants were exempted from criminal punishments and 2,033 actions associated with these cases. Here we define importance score for each action as action frequency in $C''$ divided by action frequency in $\mathbf{M}.battery$. If an action has a high importance score, this action is highly exclusive to $C''$. High exclusiveness can also lead to error actions that had very low frequency in both $C''$ and $\mathbf{M}.battery$. So we filter out 5\% most frequent actions and select 20 most important actions according to importance score. Figure \ref{fig:heatmap2} is the co-occurrences heat map of damages and selected actions. This heat map shows that fulfilling the terms in settlements for minor injuries before trial is a key factor for receiving exemption. 
    
\section{Conclusion and Future Work}
\label{sec:conclusion}
In this paper, we have sketched our approach to developing a knowledge-base to answer landscape questions revealed by judicial decision documents from Chinese courts. One of the challenges we have partially addressed centers around knowledge representation. While the current practice is to develop large scale knowledge graphs, to represent diverse data about entities, we have opted to use heterogeneous relation, a distribution matrix and a mapping between them as our knowledge structure, and showed its usefulness in answering questions. Yet, our representation has taken some simplifying decisions that failed to capture some of the practical nuances of criminal law. For example, we ``linearized" the \texttt{punishment} and \texttt{damages} dimension, while there is a variety of punishment (e.g., imprisonment and confiscation) and damage (bodily injury and financial loss) types that do not belong to the same axis. In contrast, some of the actions that have been considered as independent, should in fact be ``lumped'' into a single element. In future work, we will refine our representation to accommodate further levels of granularity.

\noindent \textbf{Acknowledgment.} This research was partially supported by the NSF RIDIR grant 1738411.
\bibliography{jurix2019}
\bibliographystyle{abbrv}
\end{document}